\documentclass[preprint,amsmath,amssymb,nofootinbib]{revtex4-1}

\usepackage{graphicx}
\usepackage{slashed}
\usepackage{bigints}

\newcommand{\ketvac}{|0\rangle}
\newcommand{\bravac}{\langle0|}
\newcommand{\vacexpect}[1]{\bravac #1\ketvac}
\newcommand{\cpp}[1]{(c^{#1})^p_p}
\newcommand{\cpgamma}[1]{(c^{#1})^p_\gamma}
\newcommand{\params}[1]{\left[#1\right]}

\begin{document}

\title{The K\"all\'en-Lehmann representation for Lorentz-violating field theory}

\author{Robertus Potting}
\affiliation{CENTRA, Departamento de F\'\i sica, 
FCT, Universidade do Algarve, 
8005-139 Faro, Portugal}

\begin{abstract}
We consider field-theoretic models, one consisting purely of scalars, the other also
involving fermions, that couple to a set of constant background coupling coefficients
transforming as a symmetric observer Lorentz two-tensor. 
We show that the exact propagators can be cast in the form of a
K\"all\'en-Lehmann representation.
We work out the resulting form of the Feynman propagators and the equal-time field
commutators, and derive sum rules for the spectral density functions.
\end{abstract}


\maketitle

\section{Introduction}
The possibility that Lorentz symmetry might be violated in Nature has received
increasing attention over the course of the last two decades or so.
Within the context of quantum field theory,
small Lorentz-violating effects can be incorporated by including
terms that contract fields and its derivatives with constant Lorentz-tensor
coefficients assumed to reside in the background.
The most general extension of the standard model of particle
physics of this type that maintains consistency conditions such as 
SU(3)$\times$SU(2)$\times$U(1) gauge invariance has been carried out
in the form of the Standard-Model Extension (SME) \cite{Colladay:1998fq}.
It allows a systematic study of and experimental searches for possible nonzero
values of the Lorentz-violating coefficients.
While most of the relevant literature considered the SME at the level of
the classical action or at tree level,
a number of studies have been done at one loop and beyond,
in particular some addressing renormalizability
\cite{one-loop-QED,nonpolynomial,one-loop-SME,renorm-Anselmi,Yukawa}.
Also causality and stability have been addressed
\cite{stability-causality,causality-CS}.
A large number of experiments have been performed to test the SME,
constraining many of its coefficients \cite{experiments}.

Lorentz-violating effects in low-energy field theory might arise from
some underlying fundamental theory,
for instance through spacetime discretization
in quantum gravity \cite{Gambini:1998it},
by spontaneous Lorentz breaking in string theory \cite{Kostelecky:1988zi},
or if spacetime is noncommutative \cite{Carroll:2001ws}.

In this work we take a step toward extending a standard result of field theory,
the K\"allen-Lehmann representation of the propagator \cite{KallenLehmann},
to SME-type actions.
We will consider two types of actions with SME-type Lorentz-violating terms.

\section{A scalar field model}
The first example we consider is the scalar-field action
\begin{equation}
S_1=\int d^4x\frac{1}{2}\left((\partial_\mu\phi)^2
+c^{\mu\nu}\partial_\mu\phi\partial_\nu\phi-m^2\phi^2\right)
\label{action-1-scalar}
\end{equation}
with $c^{\mu\nu}$ a set of constant background coupling coefficients
transforming as a symmetric traceless rank-two observer Lorentz tensor.
It should be viewed as being associated to the vacuum, its values depending on the
observer.
To indicate the $c^{\mu\nu}$ dependence of the vacuum, let us use the notation
$|0\rangle_{c^{\mu\nu}}$.
The vacuum is translation invariant:
\begin{equation}
P_\mu|0\rangle_{c^{\mu\nu}}=0,
\end{equation}
while under (observer) Lorentz transformations $\Lambda$ it transforms as
$|0\rangle_{c^{\mu\nu}}\to|0\rangle_{c'{}^{\mu\nu}}$,
with the coefficients $\{c_{\mu\nu}\}$ transforming as
\begin{equation}
c'{}^{\mu\nu}=\Lambda^\mu{}_\alpha \Lambda^\nu{}_\beta c^{\alpha\beta}.
\label{c-transformation}
\end{equation}

It is easy to check that for Lagrangian $\mathcal{L}_1$ the Lorentz-violating
coefficient $c^{\mu\nu}$ can be transformed away by an appropriate linear
transformation of the space-time variable $x^\mu$
\cite{transform-away}.
This means that the apparent Lorentz violation is an artificial one,
due to an inappropriate choice of coordinates.
It is easy, however,
to construct slightly more complicated systems in which the Lorentz
violation cannot be transformed away.
For example, we can introduce an additional scalar field and consider the Lagrangian
\begin{equation}
S_2=\int d^4x\frac{1}{2}\Bigl((\partial_\mu\phi_1)^2
+c^{\mu\nu}\partial_\mu\phi_1\partial_\nu\phi_1+(\partial_\mu\phi_2)^2 
-m_1^2\phi_1^2-m_2^2\phi_2^2-V(\phi_1,\phi_2)
\Bigr)
\label{action-2-scalar}
\end{equation}
in which only one of the two interacting fields have a direct coupling to the Lorentz-violating
coefficient.
The interaction potential $V(\phi_1,\phi_2)$ introduces radiative corrections to the self-energies
of the fields which can involve the Lorentz-violating parameter.
Particle decays may also be possible in some region of phase space, as we will discuss below. 
We will assume that $V$ is such that Lagrangian (\ref{action-2-scalar}) is renormalizable,
with Hamiltonian bounded from below, assuring stability.

Let us now consider the two-point function
\begin{equation}
D(x-y)=\vacexpect{\phi(x)\phi(y)}\,,
\vspace {3pt}
\end{equation}
where we denote $\phi(x)\equiv\phi_1(x)$ and from here on we suppress the index
$c^{\mu\nu}$ of the vacuum state. 
Writing $\phi(x)=e^{-iP\cdot x}\phi(0)e^{iP\cdot x}$ and inserting a complete set of states
$\{| n\rangle\}$ it follows
\begin{equation}
D(x-y)=\sum_n e^{-ip_n\cdot(x-y)}|\bravac\phi(0)|n\rangle|^2
\end{equation}
where $p_n^\mu$ is the momentum of the state $|n\rangle$.
If we now insert the identity
\begin{equation}
1=\int ds\int d^4p\,\theta(p^0)\delta(p^2-s)\delta^{(4)}(p-p_n)
\label{identity}
\end{equation}
it follows
\begin{eqnarray}
D(x-y)&=&\int ds\int d^4p\,\theta(p^0)\delta(p^2-s)e^{-ip\cdot(x-y)}\nonumber\\
&&\quad{}\times
\sum_n \delta^{(4)}(p-p_n)|\bravac\phi(0)|n\rangle|^2.
\label{D}
\end{eqnarray}
In usual Lorentz-invariant physics the momenta $p_n$ of the states $|n\rangle$
reside in the forward light-cone,
which limits the integrals over both $s$ and $p^0$ over positive values only.
In the presence of $c^{\mu\nu}$ couplings spacelike $p_n$  are
not guaranteed to be excluded anymore due to modifications to the dispersion
relations.
As a consequence, the integral over $s$ in (\ref{identity}) may have to
include negative values as well (for a subdomain of the three-momentum).
Here and in the following, the integral over $s$ is always assumed to be from
$-\infty$ to $\infty$, unless indicated otherwise.
The presence of states with spacelike momenta implies also that for certain
observers
the separation between the positive and negative energy branches of the dispersion
relation of certain states cannot be taken to be $p^0=0$ for all $\vec p$ anymore,
but becomes some tilted plane.
While it might be possible to accommodate such a situation
by changing the integration domain accordingly, we will, for simplicity,
assume that this does not happen for the observer frame considered here.
In particular, such a situation should not occur
when Lorentz-violating effects are small,
in the co-called ``concordant frames'' \cite{stability-causality}.

Now the quantity
\begin{equation}
\rho/(2\pi)^3\equiv
\sum_n \delta^{(4)}(p-p_n)|\bravac\phi(0)|n\rangle|^2
\label{scalar-rho}
\end{equation}
is a real, non-negative observer Lorentz scalar.
In the Lorentz-invariant case it can only depend on $\theta(p^0)$ and on $p^2$,
which is the only scalar
that can be built from the vector $p^\mu$ and the Minkowski metric $\eta_{\mu\nu}$.
In the present case,
we have an additional object that transforms as a Lorentz 2-tensor,
namely $c^{\mu\nu}$.
This means that $\rho$ can depend not just on $p^2$,
but also on $\cpp{n}$, $n=1,2,3$, where
\begin{equation}
p\cdot c^n\cdot p = p^\mu\eta_{\mu\alpha_1}c^{\alpha_1\beta_1}\eta_{\beta_1\alpha_2}
c^{\alpha_2\beta_2}\ldots\eta_{\beta_n\nu}p^\nu\equiv\cpp{n}.
\label{pcp}
\end{equation}
It is not necessary to go beyond $n=3$ as $(c^n)_{\mu\nu}$ for $n\ge4$
can be expressed as a linear combination of the quantities $(c^m)_{\mu\nu}$,
$0\le m\le3$ \cite{four-independent}.
We conclude that the quantity $\rho$ in (\ref{scalar-rho})
is a function of four independent observer scalars:%
\footnote{For certain particular forms of the Lorentz-violating tensor $c^{\mu\nu}$
it can happen that there are less than four independent observer scalars.
In particular,
this occurs if the matrix $c^{\mu\alpha}\eta_{\alpha\nu}$ has coinciding eigenvalues.
}
\begin{equation}
\rho=\rho\params{p^2,c^p_p,\cpp{2},\cpp{3}\,}\equiv \rho\params{p^2;\cpp{i}\,}.
\end{equation}
Thus the two-point function in (\ref{D}) can be written as
\begin{equation}
D(x-y)=\int ds\int \frac{d^4p}{(2\pi)^3}\theta(p^0)\delta(p^2-s)
e^{-ip\cdot(x-y)}\rho\params{s;\cpp{i}\,}. 
\label{D2}
\end{equation}

From (\ref{D2}) we find the following generalization of the Feynman propagator:
\begin{align}
D_F(x-y)&=\bravac T\phi(x)\phi(y)\ketvac\nonumber\\
&=\theta(x^0-y^0)D(x-y)+\theta(y^0-x^0)D(y-x)\nonumber\\
&=\int ds\int\frac{d^4p}{(2\pi)^3}\,\theta(p^0)\delta(p^2-s)
\rho\params{s;\cpp{i}\,}\nonumber\\
&\qquad\qquad{}\times
\bigl(\theta(x^0-y^0) e^{-ip\cdot(x-y)}+\theta(y^0-x^0)e^{ip\cdot(x-y)}\bigr).
\label{scalar-propagator}
\end{align}
The theta- and delta-functions can be traded for a momentum pole with the usual $i\epsilon$
prescription:
\begin{equation}
D_F(x-y)=\int ds\int \frac{d^4p}{(2\pi)^4}\,
\frac{i\,\rho\params{s;\cpp{i}\,}}{p^2-s+i\epsilon}\,e^{-ip\cdot(x-y)}
\label{scalar-propagator2}
\end{equation}

As to the form of the spectral density $\rho\params{s;\cpp{i}\,}$,
we expect that it can be split into contributions of a stable one-particle state
(if present, which we will suppose to be the case)
and a continuum of multi-particle states:
\begin{equation}
\rho\params{s;\cpp{i}\,}=\rho^{\mbox{\tiny 1-part}}\params{s;\cpp{i}\,}+\sigma\params{s;\cpp{i}\,}
\label{rho-1part-sigma}
\end{equation}
where $\sigma\params{s;\cpp{i}\,}$ represents the multi-particle states.
The one-particle state should satisfy an equation of motion expressing the
time derivative of the field in terms of the space derivatives.
In Fourier space, this amounts to a mass shell defining $p^0$ in terms of $\vec p$.
Consequently, we expect for $\rho^{\mbox{\tiny 1-part}}\params{s;\cpp{i}\,}$ the form
\begin{equation}
\rho^{\mbox{\tiny 1-part}}\params{s;\cpp{i}\,}=f\params{\cpp{i}\,}\delta\left(s-g\params{\cpp{i}\,}\right).
\label{rho-1part}
\end{equation}
Note that $f$ does not need to depend explicitly on $s$ due to the delta-function.
In some cases the mass-shell condition might not be expressible
as one single-valued expression for $s$ in terms of the $\cpp{i}$:
a sum of delta-function terms on the right-hand side
of (\ref{rho-1part}) might be required for a subrange of values of the $\cpp{i}$,
or some subrange might be excluded.
For simplicity, we will assume the one-particle mass-shell condition
can be expressed by (\ref{rho-1part}).
It is often useful to express $f$ and $g$ explicitly as a Taylor series:
\begin{align}
f\params{\cpp{i}\,}&=Z\Bigl(1+\hskip-6pt\sum_{1\le i_1\le\ldots i_r\le3}\hskip-6pt
f_{i_1\ldots i_r}\prod_{j=1}^r\cpp{i_j}\Bigr)
\label{f}\\
g\params{\cpp{i}}&=m^2-\hskip-6pt\sum_{1\le i_1\le\ldots i_r\le3}\hskip-6pt
g_{i_1\ldots i_r}\prod_{j=1}^r\cpp{i_j},
\label{g}
\end{align}
generalizing the Lorentz-symmetric case $f\params{\cpp{i}\,}=Z$
and $g\params{\cpp{i}}=m^2$.
Computation of the explicit values of the coefficients
$f_{i_1\ldots i_r}$ and $g_{i_1\ldots i_r}$ in (\ref{f}) and (\ref{g})
can be done by calculation of the corrections to the particle self-energy
in perturbation theory.
The propagator now takes the K\"allen-Lehmann-like form
\begin{equation}
D_F(x-y)=i\bigintsss \frac{d^4p}{(2\pi)^4}\,e^{-ip\cdot(x-y)}
\left(\frac{f\params{\cpp{i}}}{p^2 - g\params{\cpp{i}}+i\epsilon}
+\int_{M^2}^\infty\hskip-5pt ds\,\frac{\sigma\params{s;\cpp{i}\,}}
{p^2-s+i\epsilon}\right)\,.
\label{scalar-propagator3}
\end{equation}
The lower limit $M^2$ of the integral over $s$ in the second term should
be such that it includes the whole interval for which $\sigma\params{s;\cpp{i}\,}$
has support.
This generalizes the situation in the Lorentz-invariant case,
where one just takes $M$ to be the minimum value of the sum of
the physical masses of the particles in the multi-particle states into which
the particle can decay.

It is important to point out that the mass shell condition of the one-particle
propagator is modified not just by an addition of the operator $c^p_p$
(which is what one might expect naively from the form of Lagrangian
(\ref{scalar-propagator2})),
but by terms involving arbitrary products of the scalars $\cpp{i}$.
In other words, the dispersion relation can in general be expected to be
nonpolynomial in the momentum, rather than the quadratic form that has
been considered in most of the relevant literature.

The equation of motion satisfied by the one-particle propagator can be read off
from (\ref{scalar-propagator3}).
To first order in $c^{\mu\nu}$ it reads
\begin{equation}
\left(\square+m^2+g_1c^\partial_\partial
+f_1c^\partial_\partial(\square+m^2)\right)
D_F^{\mbox{\tiny 1-part}}(x-y)=-i.Z\delta^4(x-y)\,.
\end{equation}
Here we used the notation $c^\partial_\partial\equiv c^{\mu\nu}\partial_\mu\partial_\nu$.

We can derive an interesting sum rule for the spectral density by considering
the vacuum expectation value for the field commutator.
On the one hand, we have from the canonical commutation relations that follow from
the action (\ref{action-2-scalar}):
\begin{equation}
i\tilde\eta^{0\mu}\frac{\partial}{\partial x^\mu}[\phi(x),\phi(y)]|_{x^0=y^0}
=\delta^{(3)}(\vec x-\vec y)
\label{commutator1}
\end{equation}
where we defined $\tilde\eta^{\mu\nu}=\eta^{\mu\nu}+c^{\mu\nu}$.
On the other hand, we have
\begin{eqnarray}
\hskip -12pt i\tilde\eta^{0\mu}\frac{\partial}{\partial x^\mu}\bravac[\phi(x),\phi(y)]\ketvac&=&i\tilde\eta^{0\mu}\partial_\mu\int ds\int\frac{d^4p}{(2\pi)^3}
\delta(p^2-s)\epsilon(p^0)\rho\, 
e^{-ip\cdot(x-y)} \nonumber\\
&=&\int ds\int\frac{d^4p}{(2\pi)^3}
\delta(p^2-s)\epsilon(p^0)\rho\,
\tilde\eta^{0\mu}p_\mu e^{-ip\cdot(x-y)}
\label{commutator2}
\end{eqnarray}
(suppressing the arguments of $\rho\equiv\rho\params{s;\cpp{i}\,}$).
Let us now take $x^0=y^0$ in (\ref{commutator2}), compare with (\ref{commutator1}),
and integrate $\vec x$ over all space.
With the identity $\int d^3x e^{i\vec p\cdot\vec x}=(2\pi)^3\delta^{(3)}(\vec p)$
we get
\begin{eqnarray}
1 &=&\int ds\int d^4p\,\delta(p^2-s)\epsilon(p^0)\rho\params{s;\cpp{i}\,}
\tilde\eta^{0\mu}p_\mu\delta^{(3)}(\vec p)\nonumber\\
&=&\int ds \,\rho\params{s; s(c^i)^{00}}\tilde\eta^{00}
\label{sum-rule-rho}
\end{eqnarray}
Using (\ref{rho-1part-sigma}) and (\ref{rho-1part}) this can be expressed as
\begin{equation}
\left(\tilde\eta^{00}\right)^{-1}=\int ds\,f\params{s\bigl(c^i\bigr)^{00}}
\delta\left(s-g\params{s\bigl(c^i\bigr)^{00}}\right)
+\int_{M^2}^\infty \hskip-5pt ds \,\sigma\params{s;s(c^i)^{00}}\,.
\end{equation}
For small $c^{\mu\nu}$ it follows from (\ref{sum-rule-rho}),
(\ref{f}) and (\ref{g}) that $0\le Z\le 1+\mathcal{O}(c)$,
generalizing the usual relation $0\le Z\le 1$.

Above we already commented about the possibility that spacelike momenta
be included in the support of the spectral densities.
There is another situation that may occur,
that sets the Lorentz-violating situation apart from the usual case.
Namely, it can happen that the one-particle contribution to the spectral
density is distinct and well-separated from the multi-particle continuum
contribution for a large range of momenta,
but that for a certain subrange in momentum space the two become superposed.
In that case the one-particle state ceases to correspond to a stable particle,
but becomes a resonance in the multi-particle continuum.
That unstable particles appear as resonances in the multi-particle continuum
contribution to the spectral density is familiar from the Lorentz-invariant case
\cite{unstable}.
New in the Lorentz-violating case is that the same particle can be unstable
for only a subrange of momenta, while otherwise stable.

\begin{figure}[h]
\centering
\includegraphics[width=0.44\textwidth]{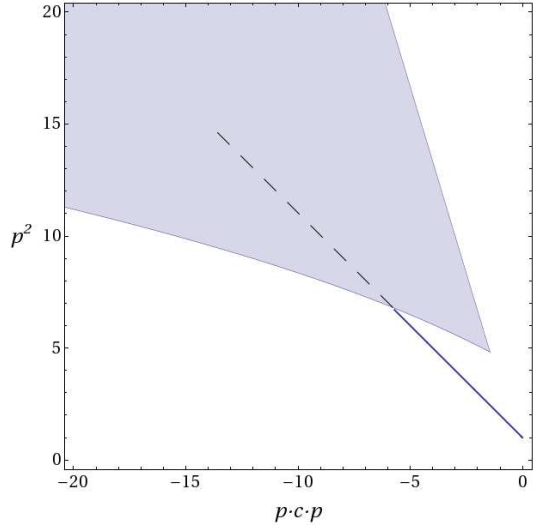}
\caption{Example of the support of a spectral density function.
The horizontal axis corresponds to $p\cdot c\cdot p$,
the vertical axis to $p^2$.
The tensor $c^{\mu\nu}$ is taken diagonal, traceless and rotationally symmetric:
$c^{00}=3c$ and $c^{ii}=c$, with $c=-1/10$.
The masses of the two scalar fields are taken equal: $m_1=m_2(\equiv1)$.
The thick line corresponds to the one-particle contribution,
the gray region to the two-particle states (type 1 + type 2).}
\label{graph}
\end{figure}
An example of such a situation is illustrated in figure~\ref{graph}.
Here we assumed the interaction potential includes a cubic term
\begin{equation}
V(\phi_1,\phi_2)\supset \lambda\phi_1^2\phi_2
\end{equation}
that may induce the Cherenkov-type decay of the (type-1) particle considered above
into the two-particle (type-1 + type-2) state $|\vec{p_1},\vec{p_2}\rangle$.
In the figure the domain of the one-particle and two-particle contributions
to the spectral densities are plotted as a function of $p^2$ and
$p\cdot c\cdot p$.
We have neglected the remaining independent observer scalars $p\cdot c^2\cdot p$
and $p\cdot c^3\cdot p$,
something that can be justified if the components of $c^{\mu\nu}$ are small.
Other multi-particle states may contribute to the spectral density as well,
but we will ignore them for simplicity.
In figure~\ref{graph} the one-particle
dispersion relation is represented by the thick line,
while the the gray region represents the two-particle states.
The latter is defined by determining all possible values that
the pair $(p\cdot c\cdot p,p^2)$ can take,
where $p^\mu$ equals the total momentum $p_1^\mu+p_2^\mu$ of the two-particle state,
with $\vec p_1$ and $\vec p_2$ the momenta of the (stable!)\ one-particle states
of type~1 and type~2, respectively.
In the figure we assumed the one-particle dispersion relations
$p_1^2+ p_1\cdot c\cdot p_1 - m_1^2=0$ and $p_2^2-m_2^2=0$ that follow from
(\ref{action-2-scalar}).
This is justified for small coupling, so that radiative corrections to the
dispersion relations can be neglected for small enough momenta
and the two-particle states can be approximated as non-interacting.

The continuation of the thick line into the gray region is indicated as well.
It does not correspond to (a continuum of) stable one-particle states anymore,
like the thick line, but rather to a continuous set of resonances
of unstable one-particle states inside the two-particle continuum.
That is, we expect the spectral density function to be strongly peaked
on the dashed line.
It is easy to show that the group velocity for the points on the dashed line
is larger than one,
as expected for Cherenkov-type decays.

In this example we have taken the tensor $c^{\mu\nu}$ to be diagonal,
traceless and rotationally symmetric:%
\footnote{This is actually an example of a special case referred to above,
in which the number of independent observer scalars is less than four
(namely, two).}
$c^{00}=3c$ and $c^{ii}=c$, with $c=-1/10$,
while the masses have been taken equal: $m_1=m_2=1$.
Note that if we take $c$ positive there is no overlap between the
one-particle state and the two-particle continuum,
amounting to a stable one-particle state for any momentum.

\section{A model with a fermion field}
As a second example let us consider as action involving a Dirac fermion
and a scalar field with a Yukawa coupling:
\begin{equation}
S_3=\int d^4x\Bigl(\bar\psi(i\partial_\mu\gamma^\mu+ic^{\mu\nu}\gamma_\mu\partial_\nu-m)\psi
+\frac12(\partial_\mu\phi)^2-\frac12(m')^2\phi^2+\lambda\phi\bar\psi\psi\Bigr)
\label{fermion-action}
\end{equation}
with $c^{\mu\nu}$ as above.
For the fermion two-point amplitude
$S(x-y)=\vacexpect{\psi_\alpha(x)\bar\psi_\beta(y)}$
we can write, analogously to (\ref{D}),
\begin{equation}
S(x-y)=\sum_n e^{-ip_n\cdot(x-y)}
\vacexpect{\psi_\alpha(0)|n\rangle\langle n|\bar\psi_\beta(0)}.
\label{fermion-twopoint}
\end{equation}
Inserting identity (\ref{identity}) this can be written as
\begin{equation}
\int ds\int d^4p\,\theta(p^0)\delta(p^2-s)e^{-ip\cdot(x-y)}
\sum_n\delta^{(4)}(p-p_n)
\vacexpect{\psi_\alpha(0)|n\rangle\langle n|\bar\psi_\beta(0)}\,.
\end{equation}
The quantity
\begin{equation}
Q_{\alpha\beta}\equiv \sum_n\delta^{(4)}(p-p_n)
\vacexpect{\psi_\alpha(0)|n\rangle\langle n|\bar\psi_\beta(0)}
\label{Q}
\end{equation}
transforms as an (observer) Lorentz scalar built out of
$\eta_{\mu\nu}$, $p^\mu$, $c^{\mu\nu}$ and $\gamma^\mu$,
and satisfies the hermiticity condition $\gamma^0Q^\dagger\gamma^0=Q$.
It follows that it can be written in the form
\begin{equation}
Q=(2\pi)^{-3}\Bigl[\rho_0 + \sum_{k=0}^3\cpgamma{k} \rho_{k+1}\Bigr]
\label{fermion-spectral-density}
\end{equation}
where $\cpgamma{k}=\gamma\cdot c^k\cdot p$ and
\begin{equation}
\rho_k\equiv\rho_k\params{p^2;c_p^p,\cpp{2}\cpp{3}\,}
\qquad (k=0,\ldots,4)
\label{fermion-rho}
\end{equation}
are real functions of the scalars $\cpp{k}$.
Eq.\ (\ref{fermion-spectral-density}) generalizes the conventional
Lorentz-invariant case that involves only $\rho_0$ and $\rho_1$.
Left out of (\ref{fermion-spectral-density}) are terms built with
$\gamma^5$, $\sigma^{\mu\nu}=i[\gamma^\mu,\gamma^\nu]/2$ and
$\gamma^5\gamma^\mu$.
The absence of the latter two follows by subjecting (\ref{fermion-twopoint})
to a combination of (observer) parity ($\mathcal{P}$)
and time reversal ($\mathcal{T}$) transformations.
From the transformation rule (\ref{c-transformation}) we find that the vacuum
$|0\rangle_{c^{\mu\nu}}$ is invariant under $\mathcal{PT}$,
which in turn implies invariance of  $S(x-y)$.
This precludes the presence of $\sigma^{\mu\nu}$
and $\gamma^5\gamma^\mu$ in (\ref{fermion-spectral-density}),
as they transform with the wrong sign.
The absence of a term proportional to $\gamma^5$ can be argued by
applying only $\mathcal{P}$ to $S(x-y)$ instead.
In this case the vacuum is no longer invariant, as $c^{\mu\nu}$ transforms
into $(-)^\mu(-)^\nu c^{\mu\nu}$ (where $(-)^\mu$ equals $1$ or $-1$ for
the time and space components, respectively).
If we insist that (\ref{fermion-spectral-density}) be form invariant,
a term proportional to $\gamma^5$ is precluded.

Of prime interest is again the time-ordered product:
\begin{align}
S_F(x-y)_{\alpha\beta}&=\vacexpect{T\psi_\alpha(x)\bar\psi_\beta(y)}\nonumber\\
&=\theta(x^0-y^0)S(x-y)_{\alpha\beta}-\theta(y^0-x^0)\bar S(x-y)_{\alpha\beta}
\label{fermion-propagator}
\end{align}
Here the reverse-field amplitude in the second term
\begin{equation}
\bar S(x-y)_{\alpha\beta}=\vacexpect{\bar\psi_\beta(y)\psi_\alpha(x)}
\end{equation}
can be found by applying the (unitary) charge conjugation operation $\mathcal{C}$
to the first term.
Applying the defining relations
\begin{align}
\psi(x)&\to \mathcal{C}\psi(x)\mathcal{C}^\dagger=C\bar\psi^T(x),
\label{Cpsi}\\
\bar\psi(x)&\to  \mathcal{C}\bar\psi(x)\mathcal{C}^\dagger=-\psi^T(x)C^T,
\label{Cbarpsi}
\end{align}
with the charge conjugation matrix $C$ satisfying
(in the Dirac representation \cite{Itzykson:1980rh})
$C=i\gamma^0\gamma^2$, $C^T=C^{-1}=-C$,
it follows that
\begin{eqnarray}
\bar S(x-y)_{\alpha\beta}
&=&\vacexpect{\mathcal{C}^\dagger\mathcal{C}\bar\psi_\beta(y)\psi_\alpha(x)\mathcal{C}^\dagger\mathcal{C}}\nonumber\\
&=&\vacexpect{\mathcal{C}\bar\psi_\beta(y)\mathcal{C}^\dagger\mathcal{C}\psi_\alpha(x)\mathcal{C}^\dagger}\nonumber\\
&=&-\bigl(C_{\alpha\sigma}\vacexpect{\psi_\sigma(y)\bar\psi_\tau(x)}C^T_{\tau\beta}\bigr)\nonumber\\
&=&-\bigl(C.S(x-y).C^T\bigr)^T_{\alpha\beta}
\label{C-twopoint}
\end{eqnarray}
In the second equality we used that the vacuum is invariant under charge conjugation,
which continues to be the case even in the presence of $c^{\mu\nu}$.

Using (\ref{C-twopoint}) and the equality
$\bigl(C\gamma^\mu C^T\bigr)=-\bigl(\gamma^\mu\bigr)^T$
we obtain for the fermion propagator (\ref{fermion-propagator}):
\begin{align}
S_F(x-y)=&\int ds\int 
\frac{d^4p}{(2\pi)^3}\theta(p^0)\delta(p^2-s)\nonumber\\
&{}\times\biggl[\theta(x^0-y^0)e^{-ip\cdot(x-y)}
\Bigl(\rho_0+\sum_{k=0}^3\cpgamma{k}\,\rho_{k+1}\Bigr)\nonumber\\
&\qquad{}+\theta(y^0-x^0)e^{ip\cdot(x-y)}
\Bigl(\rho_0-\sum_{k=0}^3\cpgamma{k}\, \rho_{k+1}\Bigr)\biggr]\nonumber\\
=&\int ds\int\frac{d^4p}{(2\pi)^4}\frac{i.e^{-ip\cdot(x-y)}}{p^2-s+i\epsilon}
\Bigl(\rho_0+\sum_{k=0}^3\cpgamma{k}\,\rho_{k+1}\Bigr).
\label{fermion-propagator2}
\end{align}
In the last equality we used the identity
\begin{align}
\int \frac{d^4p}{(2\pi)^4}\frac{e^{-ip\cdot(x-y)}f(p^\mu)}{p^2-s+i\epsilon}
=&i\int \frac{d^4p}{(2\pi)^3}\theta(p^0)\delta(p^2-s)\nonumber\\
&{}\quad\times\Bigl[\theta(x^0-y^0) e^{ip\cdot(x-y)}f(-p^\mu)
+\theta(y^0-x^0)e^{ip\cdot(y-x)}f(p^\mu)\Bigr].
\end{align}
Supposing again, as in the scalar case,
that the spectrum includes a stable one-particle state,
we expect to be able to identify a one-particle pole and a continuum
contribution of multi-particle states:
\begin{equation}
\rho_k\params{s;\cpp{j}\,}=\rho_k^{\mbox{\tiny 1-part}}\params{s;\cpp{k}\,}
+\sigma_k\params{s;\cpp{k}\,}\,.
\label{rhoi-1i-sigmai}
\end{equation}
Analogously to the scalar case (\ref{rho-1part}),
$\rho_k^{\mbox{\tiny 1-part}}$ is proportional to a delta-function of the mass shell condition:
\begin{equation}
\rho^{\mbox{\tiny 1-part}}_k\params{s;\cpp{i}\,} =
f_k\params{\cpp{i}\,}\delta\left(s-g\params{\cpp{i}\,}\right)\,.
\label{rho-1part-fermion}
\end{equation}
Note that we take the mass-shell condition for different values
of the index $k$ to coincide;
however, the functions $f_k\params{\cpp{i}\,}$ can depend on $k$.
The function $g\params{\cpp{i}\,}$ can be expanded as in (\ref{g}),
while for the $f_k$ we get analogously to (\ref{f}):
\begin{equation}
f_k\params{\cpp{i}\,}=Z_k\Bigl(1+\hskip-6pt\sum_{1\le i_1\le\ldots i_r\le3}\hskip-6pt
f_{k,i_1\ldots i_r}\prod_{j=1}^r\cpp{i_j}\Bigr)
\label{fk}
\end{equation}
The fermion propagator $S_F(x-y)$ becomes
\begin{align}
i\bigints \frac{d^4p}{(2\pi)^4}\,e^{-ip\cdot(x-y)}
&\left(\frac{\displaystyle f_0\params{\cpp{i}}+\sum_{k=0}^3 f_{k+1}\params{\cpp{i}}\cpgamma{k}}
{\displaystyle p^2+\hskip-5pt\sum_{1\le i_1\le\ldots i_r\le3}\hskip-5pt
g_{i_1\ldots i_r}\prod_{j=1}^r\params{\cpp{i_j}}-m^2+i\epsilon}\right.\nonumber\\
&\qquad{}+\bigints_{M^2}^\infty\hskip-5pt ds\>
\left.\frac{\displaystyle\sigma_0\params{s;\cpp{i}\,}+\sum_{k=0}^3 \sigma_{k+1}\params{s;\cpp{i}\,}\cpgamma{k}}
{\vphantom{\displaystyle\prod_{j=1}^r} p^2-s+i\epsilon}
\>\>\right)
\label{fermion-propagator3}
\end{align}
The first term, the one-particle contribution to the propagator,
is a generalization of the Lorentz-invariant case,
in which only the coefficients $f_0$ and $f_1$ contribute.
We expect that the one-particle propagator be the Green's function of a Dirac-type
operator of the (momentum-space) form $\alpha_0+\sum_{k=0}^3\cpgamma{k}\alpha_{k+1}$,
with coefficients $\alpha_k$ possibly depending on $\cpp{i}$ $(i>0)$.
For this to be the case, we need the identity
\begin{equation}
\left(\alpha_0+\sum_{k=0}^3\cpgamma{k}\alpha_{k+1}\right)
\left(\frac{f_0+\sum_{k=0}^3\cpgamma{k}f_{k+1}}
{p^2-g\params{\cpp{i}}}\right)=1
\label{identity-fermion-greensfunction}
\end{equation}
(with $f_k\equiv f_k\params{\cpp{i}}$ and $\alpha_k=\alpha_k\params{\cpp{i}}$)
to hold identically.
This is solved by the ansatz $\alpha_0=\alpha.f_0$, $\alpha_k=-\alpha.f_k$ $(k>0)$,
with $\alpha=\alpha\params{\cpp{i}}$,
provided $\alpha$, $f_k$ and $g$ satisfy the relation
\begin{equation}
\alpha\left(f_0^2-\sum_{k,l\ge1}f_k f_l\cpp{k+l-2}\right)=p^2-g\params{\cpp{i}}.
\label{identity-fermion-greensfunction2}
\end{equation}
Here the quantities $\cpp{i}$ with $i\ge4$ should be expressed as a linear combination
of $\cpp{j}$ with $0\le j\le3$, as explained below eq.\ (\ref{pcp}).
By matching the coefficients of $p^2$ on both sides of eq.\ (\ref{identity-fermion-greensfunction2})
it follows that $\alpha=-(f_1)^{-2}+\mathcal{O}(c^4)$,
while the coefficients $g_{i_1\ldots i_r}$ can be expressed in terms of the $f_{k,i_1\ldots i_r}$.
Working out the $\alpha_i$ to first order in $c^{\mu\nu}$
one finds that the one-particle fermion propagator satisfies
\begin{equation}
\left((1+f_{1,1}c^\partial_\partial)i\slashed\partial-m\bigl(1-(f_{0,1}-2f_{1,1})c^\partial_\partial\bigr)
+\frac{Z_2}{Z_1}c^\partial_\partial\right)
S^{F,\mbox{\tiny 1-part}}(x-y)=iZ_1\delta^4(x-y).
\end{equation}

Finally, we derive some positivity relations and sum rules for the spectral densities.
First consider the quantity Tr$[\gamma^0Q]$ (see (\ref{Q})),
which is positive definite for any $p$, because
\begin{equation}
\sum_n\delta^{(4)}(p_n-p)\sum_\alpha\vacexpect{\psi_\alpha(0)|n\rangle
\langle n|\psi^\dagger_\alpha(0)}
=\sum_n\delta^{(4)}(p_n-p)\sum_\alpha|\bravac\psi_\alpha(0)|n\rangle|^2\ge0\,,
\label{TrQ1}
\end{equation}
while on the other hand we have from (\ref{fermion-spectral-density})
\begin{equation}
\mbox{Tr}[\gamma^0Q]=\frac{1}{2\pi^3}\Bigl(p^0\rho_1
+\sum_{k=1}^3\bigl(c^{\mu0}p_\mu\bigr)\rho_{k+1}\Bigr)\ge0\,.
\label{TrQ2}
\end{equation}
Note again that the $\rho_k$ depend on the $\cpp{i}$.
To zeroth order in $c^{\mu\nu}$ (\ref{TrQ2}) yields the conventional result $\rho_1\ge0$.

More generally, we can work out the positive-definite quantity
\begin{equation}
\sum_\alpha|\bravac\Bigl[\Bigl(i(\lambda_1\slashed\partial+\sum_{k=1}^3\lambda_{k+1}
(\gamma\cdot c^k\cdot\partial)-\lambda_0\Bigr)\psi(0)\Bigr]_\alpha|n\rangle|^2
\end{equation}
for arbitrary coefficients $\lambda_k$.
From this it is not difficult to derive the relation
\begin{equation}
\sum_{i,j=0}^3\cpp{i+j}\rho_{i+1}\rho_{j+1}-(\rho_0)^2\ge0\,,
\end{equation}
generalizing the conventional result $\rho_1\ge|\rho_0|/\sqrt{s}$.

Consider now the canonical equal-time anti-commutation relation that follows
from the fermion action (\ref{fermion-action}):
\begin{equation}
\bigl\{\psi_\alpha(\vec x),\bigl(\bar\psi(\vec y)\Gamma^0\bigr)_\beta\bigr\}
=\delta^{(3)}(\vec x-\vec y)\,.
\label{fermion-canonical-anticommutator}
\end{equation}
with $\Gamma^0=\tilde\eta^{0\mu}\gamma_\mu$.
By inspecting (\ref{fermion-propagator})
and (\ref{fermion-propagator2}) it follows that we can write the anticommutator also as
\begin{align}
\vacexpect{\{\psi_\alpha(x),\bar\psi_\beta(y)\}}=&
\int ds\int \frac{d^4p}{(2\pi)^3}\theta(p^0)\delta(p^2-s)\times\nonumber\\
&{}\times\biggl[e^{-ip\cdot(x-y)}
\Bigl(\rho_0+\sum_{k=0}^3\cpgamma{k}\rho_{k+1}\Bigr)
-e^{ip\cdot(x-y)}
\Bigl(\rho_0-\sum_{k=0}^3\cpgamma{k}\rho_{k+1}\Bigr)
\biggr]_{\alpha\beta}
\label{fermion-anticommutator}
\end{align}
Multiplying (\ref{fermion-anticommutator}) by $\Gamma^0$,
equating the result to (\ref{fermion-canonical-anticommutator}) and integrating both sides over all space at $x^0=y^0$ yields the result
\begin{equation}
1=\int ds\sum_{k=0}^3\rho_{k+1}
\params{s;s(c^i)^{00}}(c^k)^{\mu0}\gamma_\mu\Gamma^0.
\label{fermion-sum-rule}
\end{equation}
Evaluating this to order $\mathcal{O}(c^1)$ and matching both sides yields
the relations
\begin{eqnarray}
1&=&\frac{1+f_{1,1}c^{00}}{1+g_1c^{00}}Z_1+\int_{M^2}^\infty \sigma_1\params{s;\bigl(c^i\bigr)^{00}}\,,
\label{fermion-sum-rule-1}\\
-1&=&Z_2+\int_{M^2}^\infty \sigma_2\params{s;\bigl(c^i\bigr)^{00}}\,.
\label{fermion-sum-rule-2}
\end{eqnarray}
Here (\ref{fermion-sum-rule-1}) is correct up to $\mathcal{O}(c)$,
(\ref{fermion-sum-rule-2}) up to $\mathcal{O}(c^0)$.
Evaluating (\ref{fermion-sum-rule}) to higher order in $c^{\mu\nu}$ yields sum rules
for $Z_3$ and $Z_4$ as well.
From (\ref{fermion-sum-rule-1}) we conclude $0\le Z_1\le 1+(g_1-f_{1,1})c^{00}+\mathcal{O}(c^2)$.

\section{Conclusions}
We showed how a K\"all\'en-Lehmann type representation can be
derived for field theory models with scalars and/or fermions
that include SME-type Lorentz-violating parameters.
Two specific models were considered,
one involving only scalar fields,
the other including also a fermion field,
coupled to a $c^{\mu\nu}$ tensor coefficient that violates (particle)
Lorentz invariance.
It will be interesting to extend our result to the full Standard-Model Extension,
that includes all possible Lorentz-violating tensor couplings,
as well as gauge fields (which we have not considered here).
Introducing additional Lorentz-violating parameters, while straightforward,
will complicate the most general form that the spectral density functions
(\ref{scalar-rho}) and (\ref{fermion-rho}) can take,
as they will increase the number of independent 
observer scalars on which the spectral densities can depend.

This work constitutes one of the first
truly nonperturbative results derived in the context of Lorentz-violating
field theory.
One of the key ideas of this work is that the spectral densities in the
presence of Lorentz violation depend on momentum-dependent observer
scalars other than just $p^2$.
Their general structure was worked out and used to write down the
explicit form of the exact propagators,
exhibiting the one-particle and multi-particle contributions.

It will be of great interest to work out physical consequences of the
Kallen-Lehmann representation in these Lorentz-violating models,
such as an analysis of the fate of particles that are unstable for
a subrange of momenta.

\vspace*{5mm}

\begin{acknowledgments}
It is a pleasure to thank Alan Kosteleck\'y and Ralf Lehnert
for suggestions and discussions.
Financial support by the Portuguese Funda\c c\~ao para a Ci\^encia e a Tecnologia
is gratefully acknowledged.
\end{acknowledgments}

\end{document}